%%%%%%%%%%%%%%%%%%%%%%%%%%%%%%%%%%%%%%%%%%%%%%%%%%%%%%%%%%%%%%%%
%%%%%%%%  Parafermionic derivation of Andrews-type multiple sums
%%%%%%%%  P. Jacob and P. Mathieu
%%%%%%%%%%%%%%%%%%%%%%%%%%%%%%%%%%%%%%%%%%%%%%%%%%%%%%%%%%%%%%%%

%\magnification1100

%%%%%%%%%%%%%%%% pour tester la longueur de LMP%%%%%%%%
% \setlength{\textwidth}{130mm}
%  \setlength{\textheight}{198mm}
 % \usepackage{mathptm}
%%%%%%%%%%%%%%%%%%%%%%%%%%%%%%%%%%%%%%%%%%%%%%%%%%%%%%%%

\input harvmac.tex

%\draft

%-------------------------------------------------------------------------------
% contractions de Wick
%
\def\ubrackfill#1{$\mathsurround=0pt
	\kern2.5pt\vrule depth#1\leaders\hrule\hfill\vrule depth#1\kern2.5pt$}
\def\contract#1{\mathop{\vbox{\ialign{##\crcr\noalign{\kern3pt}
	\ubrackfill{3pt}\crcr\noalign{\kern3pt\nointerlineskip}
	$\hfil\displaystyle{#1}\hfil$\crcr}}}\limits
}

\def\ubrack#1{$\mathsurround=0pt
	\vrule depth#1\leaders\hrule\hfill\vrule depth#1$}
\def\dbrack#1{$\mathsurround=0pt
	\vrule height#1\leaders\hrule\hfill\vrule height#1$}
\def\ucontract#1#2{\mathop{\vbox{\ialign{##\crcr\noalign{\kern 4pt}
	\ubrack{#2}\crcr\noalign{\kern 4pt\nointerlineskip}
	$\hskip #1\relax$\crcr}}}\limits
}
\def\dcontract#1#2{\mathop{\vbox{\ialign{##\crcr
	$\hskip #1\relax$\crcr\noalign{\kern0pt}
	\dbrack{#2}\crcr\noalign{\kern0pt\nointerlineskip}
	}}}\limits
}

\def\ucont#1#2#3{^{\kern-#3\ucontract{#1}{#2}\kern #3\kern-#1}}
\def\dcont#1#2#3{_{\kern-#3\dcontract{#1}{#2}\kern #3\kern-#1}}

%=================================================================
% MACROS

%* on my mac I do not have these: so start by decommenting
%* the following set and comment mine

%\input amssym.def \def\Z{{\Bbb Z}} \def\R{{\Bbb R}} \def\Q{{\Bbb Q}}
%\def\C{{\Bbb C}} \def\M{{\Bbb M}} \def\N{{\Bbb N}}
%added this macro
%\def\Z{{\Bbb Z}}
%\def\QQ{{\Bbb Q}}

 % TABLEAU NOIR (BLACKBOARD)
\font\tenmsy=msbm10
\font\sevenmsy=msbm10 at 7pt
\font\fivemsy=msbm10 at 5pt
\newfam\msyfam % family 11
\textfont\msyfam=\tenmsy
\scriptfont\msyfam=\sevenmsy
\scriptscriptfont\msyfam=\fivemsy
\def\blackB{\fam\msyfam\tenmsy}
\def\Z{{\blackB Z}}

% TABLEAU NOIR (BLACKBOARD)
\font\tenmsy=msbm10
\font\sevenmsy=msbm10 at 7pt
\font\fivemsy=msbm10 at 5pt
\newfam\msyfam % family 11
\textfont\msyfam=\tenmsy
\scriptfont\msyfam=\sevenmsy
\scriptscriptfont\msyfam=\fivemsy
\def\blackB{\fam\msyfam\tenmsy}
\def\M{{\cal M}}

\let\d\partial

\let\s\sigma

\let\R\rangle

\let\vph\varphi

\def\frac#1#2{{\textstyle{#1\over #2}}}

% alignements multiples

\def\text#1{\quad\hbox{#1}\quad}

\def\e{\epsilon}

\def\A{{\cal{A}}}

\def\y{{\infty}}

\def\rw{\rightarrow}

\def\R{\rangle}

\def\osp{\widehat{osp}}

%\def\E1{E_{1}}

% Equations (overrides harvmac's equation macros)
\newcount\eqnum
\eqnum=0
\def\eq{\eqno(\secsym\the\meqno)\global\advance\meqno by1}
\def\eqlabel#1{{\xdef#1{\secsym\the\meqno}}\eq }

% References (overrides harvmac's reference macros)
\newwrite\refs
\def\startreferences{
  \immediate\openout\refs=references
  \immediate\write\refs{\baselineskip=14pt \parindent=16pt \parskip=2pt}
}
\startreferences

\refno=0
\def\aref#1{\global\advance\refno by1
  \immediate\write\refs{\noexpand\item{\the\refno.}#1\hfil\par}}
\def\ref#1{\aref{#1}\the\refno}
\def\refname#1{\xdef#1{\the\refno}}

\newcount\exno
\exno=0
\def\Ex{\global\advance\exno by1{\noindent\sl Example \the\exno:

\nobreak\par\nobreak}}

\parskip=6pt

\overfullrule=0mm
%%%%%%%%%%%%%%%%%%%%%%%%%%%%
\def\frac#1#2{{#1 \over #2}}

\let\d=\partial

\def\suh{{\widehat {su}}}
\def\uh{{\widehat u}}
\def\rw{{\rightarrow}}

% References (overrides harvmac's reference macros)
\newwrite\refs
\def\startreferences{
  \immediate\openout\refs=references
  \immediate\write\refs{\baselineskip=14pt \parindent=16pt \parskip=2pt}}
\startreferences \refno=0
\def\aref#1{\global\advance\refno by1
  \immediate\write\refs{\noexpand\item{\the\refno.}#1\hfil\par}}
\def\ref#1{\aref{#1}\the\refno}
\def\refname#1{\xdef#1{\the\refno}}

%===============================================================================

% \Title{\vbox{\baselineskip12pt
 \hbox{DCPT-05/19}
%}\hbox{LETH-PHY-2/92}\hbox{hepth@xxx/9203004}}}
% {\vbox {\centerline{Can fusion coefficients be calculated}
% \bigskip
% \centerline{from the depth rule ?}
% }}

% PAGE TITRE
\Title{\vbox{\baselineskip12pt
\hbox{ }}}
{\vbox {\centerline{Parafermionic derivation of Andrews-type multiple sums}
}}

\smallskip
\centerline{ P. Jacob and P. Mathieu
% \foot{Work supported by EPSRC (UK) and NSERC (Canada).}%  and FCAr (Qu\'ebec) }
% \foot{Corresponding author (PM): tel:
% 418-656-3416; fax: 418-656-2040 }
}
\smallskip\centerline{ Department of Mathematical Sciences, University of Durham, Durham, DH1 3LE, UK}
\centerline{and}
\centerline{D\'epartement de
physique, Universit\'e Laval, Qu\'ebec, Canada G1K 7P4}
\smallskip\centerline{(patrick.jacob@durham.ac.uk, pmathieu@phy.ulaval.ca)}
\vskip .2in
\bigskip
%\bigskip
%\bigskip
%\bigskip
\bigskip
\centerline{\bf Abstract}

A multi-parafermion basis of states for the $\Z_k$ parafermionic models is derived. Its generating function is  constructed by elementary steps. It corresponds to the  Andrews multiple-sum which enumerates partitions whose parts  separated by the distance $k-1$ differ by at least   2. Two  analogous bases are derived for graded parafermions; one of these entails a new expression for their fermionic characters.

%Classification numbers: 11.10.-z, 02.20+bm

%Keywords: conformal field theory, parafermions,
%$Z_N$ models; character; singular vectors.
%\Date{10/03\ }
%\ (hepth@xxx/0505087)}

\let\n\noindent
\vfill
\eject

%==============================================================================
\headline = {\hfil \folio }

\newsec{Introduction}

\subsec{Rogers-Ramanujan identities and the Andrews-Gordon generalization}

The search for fermionic-type  characters, that is, characters  expressed as positive sums, has brought the topic of Rogers-Ramanujan identities within the framework of conformal field theory [\ref{R. Kedem, T.R. Klassen, B. M. McCoy and E. Melzer, Phys. Lett. {\bf B304} (1993) 263 and  Phys. Lett. {\bf B307} (1993) 68; R. Kedem, B. M. McCoy and E. Melzer, in {\it Recent progress in statistical mechanics and quantum field theory},
ed. by P. Bouwknegt et al, World Scientific (1995) 195. }\refname\KKMM]\foot{For further early references and a brief review of fermionic-type  characters, see the introduction of [\ref{P. Jacob and
P. Mathieu,  Nucl. Phys. {\bf B 620} (2002)
351.}\refname\JMb].}. The Rogers-Ramanujan identities are
$$\sum_{m\geq 0} { q^{m^2+(2-i)m}\, z^m\over (q)_m} = \prod_{n\not=0,\pm i\; {\rm mod}\; 5}{1\over 1-q^n}\;,\qquad\qquad (i=1,2)\eqlabel\RRs$$
where
$$
(a)_n=(a;q)_n= \prod_{i=0}^{n-1} (1-aq^i)\; .
\eq$$
This has various generalizations, the most relevant one being the Andrews-Gordon identity (see e.g., [\ref{G.E. Andrews, {\it The theory of
partitions}, Cambridge Univ. Press, Cambridge, UK, (1984).}\refname\Andr]):
$$ \sum_{m_1,\cdots,m_{k-1}=0}^\y {
q^{N_1^2+\cdots+ N_{k-1}^2+N_i+\cdots +N_{k-1}} \over (q)_{m_1}\cdots (q)_{m_{k-1}}
}= \prod_{n\not=0,\pm i\; {\rm mod}\; 2k+1}{1\over 1-q^n}\;,\qquad\qquad (i=1,\cdots, k)\eqlabel\AG$$
with $N_j$ defined as
$$
 N_j= m_j+\cdots
+m_{k-1}\,.
% \qquad L_j=N_j+\cdots N_{k-1}\,,\qquad N=L_1\; ,
\eqlabel\efNL$$
% ($ L_k=L_{k+1}=0$) 
The identity (\AG) has the following combinatorial interpretation: the lhs is the generating function for 
partitions $ (n_1,n_2, \cdots)$ subject to the  difference 2 condition %between parts at distance $k-1$
% \footnote{In the following the qualitative  
% `$\ell$-step difference-$r$ condition' applied to a set 
% of integers $\{a_j\}$ refers to
% the following condition: $a_j\geq a_{j+\ell}+r$.}
$$
 n_j\geq n_{j+k-1}+2\;, 
\eqlabel\usual$$
and containing at most $i-1$ parts equal to 1, while the rhs is the generating function for partitions without parts equal to $0,\pm i\; {\rm mod}\; 2k+1$. 

In the context of conformal field theory, we are mainly interested in the lhs, which is a fermionic-type expression. Granting that the two sets of partitions just described are equinumerous (which is the Gordon identity), the difficult part in establishing the analytic version (\AG) of this combinatorial identity is to demonstrate that the lhs is the proper generating function for partitions restricted by (\usual).

The point of this paper is to show that conformal field theory provides a simple method for  constructing the sum-side of (\AG) and related extensions. But to put this statement in perspective, lets us turn to some remarks concerning the  Andrews multiple-sum.

\subsec{Remarks on the Andrews multiple-sum}

The generating function for
partitions $ (n_1,\cdots,n_m)$ with prescribed number of parts subject to the  difference 2 condition (\usual) and containing at most $i-1$ parts equal to 1,  is
$$
F_{k,i}(z;q)= \sum_{m_1,\cdots,m_{k-1}=0}^\y {
q^{N_1^2+\cdots+ N_{k-1}^2+N_i+\cdots +N_{k-1}}\; z^{N_1+\cdots +N_{k-1}} \over (q)_{m_1}\cdots (q)_{m_{k-1}}
}\;,\eqlabel\defF$$
where the power of $z$ gives the length of the  partition. 
The standard proof of this result is  based on the following indirect trick [\ref{G.E. Andrews, Houston J. Math. {\bf 7} (1981) 11.}]  (see also 
[\Andr]  chap 7). One first shows that the number $f_{k,i}(m,n)$ of partitions of $n=\sum n_i$ with $m$ parts subject to (\usual), and containing at most $i-1$ parts equal to 1, satisfies a simple  recurrence relation on $i$. This is then lifted to a recurrence relation for the generating function:
$$F_{k,i}(z;q)= \sum_{m,n\geq 0} z^m q^n f_{k,i}(m,n)\;. \eq$$
Finally, it is proved that the multiple sum on the right hand side of (\defF) does satisfy  this recurrence relation, with the same boundary conditions.  The uniqueness of the solution of this recurrence problem completes the proof.  But this is clearly  a verification proof and not a constructive one.\foot{The original proof is based on the same recurrence relation for the generating function but the recurrence is not yet rooted  to the restricted partitions, that is, to $f_{k,i}$ [\ref{
G.E. Andrews,  Proc. Nat. Acad. Sci. USA {\bf 71} (1974)
4082.}].}

To our knowledge, there are no elementary constructive proofs of (\defF).\foot{There are constructive proofs, using either Durfee dissections [\ref{G.E. Andrews, Am. J.  Math. {\bf 101} (1977) 735.}] Êor  a bijection to lattice paths  [\ref{D. Bressoud, {\it Lattice paths and Rogers-Ramanujan identities }, in {\it Number Theory, Madras 1987},
ed. K. Alladi. Lecture Notes in Mathematics {\bf 1395} (1987) 140.}]Ê (see also [\ref{S.O. Warnaar, Comm. Math. Phys. {\bf 184} (1997) 203.}]) but (arguably) these are not quite elementary.} To illustrate what is meant by such a proof, consider the case $k=2$.
The multiple sum reduces then to the sum-side of the Rogers-Ramanujan identity. As it is well-known,  the generating function $F_{2,i}$ is easily derived.  Take $i=2$. Looking for the generating function of partitions subject to the condition
$$n_j\geq n_{j+1}+2\;,\eq$$
one first counts those restricted partitions of length  $m$  
% (i.e., the number of non-zero parts)
 and then sum over $m$. These  restricted partitions can be described by the set of (standard)  partitions of length at most $m$, whose generating function is $(q)_m^{-1}$, to which we add the `staircase' $(2m-1,\cdots, 5,3,1)$. Since the weight of the staircase is $q^{m^2}$, we end up with the following expression for $F_{2,2}$:
$$F_{2,2}(z;q)= \sum_{m\geq 0} { q^{m^2}\, z^m\over (q)_m} \;, \eq$$
where  the variable $z$ has been introduced to keep track of the length. For $i=1$, there are no 1, so that the staircase is shifted to $(2m,\cdots, 6,4,2)$ and this  produces an extra term $q^m$ within the sum. We thus recover the generating function $F_{2,i}(z;q)$ for $i=1,2$  by elementary steps. 

How does this simple argument breaks for $k>2$? Let us take $k=3$ to illustrate the point and set $i=3$. The `ground state' that replaces the staircase of the previous example, is now $(\cdots , 7,5,5,3,3,1,1)$. To use the same  strategy as for the $k=2$ case would amount to trying  to describe all partitions of length $m$ with
$$n_j\geq n_{j+2}+2\;, \eqlabel\thre$$
in terms of the usual partitions of length at most $m$, to which we add the contribution the ground state $(\cdots , 7,5,5,3,3,1,1)$. But this simple description is simply not correct when $k>2$. This can be seen plainly from a counter-example.  There are three allowed partitions of length 3 and weight 7 satisfying (\thre): $(5,1,1)$, $(4,2,1)$ and $(3,3,1)$. Subtracting the ground-state contribution $(3,1,1)$, we are left with $(2,0,0),\ (1,1,0)$ and $(0,2,0)$. But $(0,2,0)$ is not a genuine partition. This shows neatly that the argument used for $k=2$ cannot be extended to higher value of $k$. This is a simple rationale justifying the non-elementary aspect of the proofs of  (\defF).

% In the present work, we propose an elementary
%  constructive proof of $F_{k,i}(z;q)$ based on 
% parafermionic conformal field theory. 

% To put this statement in  perspective, let us recall that

\subsec{The Andrews multiple-sum in conformal field theory}

We present here an elementary conformal-field-theoretical derivation of  $F_{k,i}(z;q)$. As already mentioned,
the multiple sum $F_{k,i}(z;q)$ has appeared in the description of the basis of states of some conformal field theories. In particular, with $z=1$, it gives the irreducible (normalized)  characters of the minimal models $\M(2,p)$ [\ref{B.L. Feigin, T. Nakanishi  and H. Ooguri,   Int. J. Mod. Phys.
{\bf A7} Suppl. {\bf 1A} (1992) 217.
}\refname\FNO]. But more important for us here is that for a different specialization of $z$, one recovers the characters of the parafermionic $\Z_k$ models in their fermionic form [\ref{J. Lepowsky and
M. Primc,  Contemporary Mathematics {\bf 46} AMS, Providence, 1985.}\refname\LP, \JMb].

But how does this function $F_{k,i}(z;q)$ actually appear  in the parafermionic context? Using the generalized commutation relations between the modes of the basic parafermionic field and implementing the $\Z_k$ invariance, we end up with a description of the basis of states formulated in terms of the condition (\usual), where parts at distance $k-1$ differ by 2. More precisely, if $\A^{(1)}$ denotes the  modes of the basic parafermionic field $\psi_1$ of dimension $1-1/k$, the descendent states are of the form\foot{Here the mode is defined up to a fractional part  that is  irrelevant for the present discussion.}
$$\A^{(1)}_{-n_1}\cdots \A^{(1)}_{-n_m}|{\rm hws}\R\;, \eq$$
with the $n_i$ being positive integers subject to (\usual) and $|{\rm hws}\R$ stands for a highest-weight state.  There is in addition a boundary condition that specifies the irreducible module (the highest-weight state) under consideration. With the module labeled by an integer $1\leq i\leq k$, this condition reads:
$$n_{m-i+1} \geq 2\;.  \eq$$ 
This is clearly equivalent to the previously mentioned condition that specifies the maximal number of 1 that can appear at the right end of the associated partition $(n_1,\cdots,n_m)$.  At this point, i.e., having reached a description of the basis of states, the generating function (\defF) is invoked. Finally, by relating $z$ to a power of $q$ in order to adjust the total power of $q$ to the proper conformal dimension of the states (taking thus due care of the omitted fractional parts in the modes), we recover the  irreducible parafermionic characters.\foot{This construction could be rephrased in more Lie-algebraic  terms in the language of vertex operator algebras following  [\LP].}

In this work, we present another basis of states for the parafermionic models. This  basis is not formulated solely in terms of the basic parafermionic modes but involves rather the modes of the complete set of $k-1$ parafermionic fields. The generating function of this basis of states turns out to be built by elementary steps, analogous to those that led to the sum-side of the Rogers-Ramanujan identity. The resulting expression is precisely the above function $F_{k,i}(z;q)$. Turning this around, the equivalence of the two bases of states for the parafermionic theories, the one exposed here and the previous one formulated in terms of partitions restricted by (\usual), entails a simple constructive proof of the Andrews multiple-sum identity.

Physically, this new derivation is quite appealing since each of the  $k-1$ sums on the rhs of (\defF) is  linked to the counting of a given type of modes. In other words, the number $m_j$ labels the number of parafermionic modes of type $j$.

\subsec{The $\Z_k$ multi-parafermion basis: combinatorial formulation}

Let us state our result in a field-theoretical independent way. The multi-parafermion basis of states is equivalent to the set of $k-1$ ordered partitions of respective lengths $m_1,\cdots , m_{k-1}$, i.e., 
$$(n^{(1)}, n^{(2)}, \cdots ,n^{(k-1)})\qquad {\rm with}\qquad 
n^{(j)}= (n^{(j)}_1, \cdots , n^{(j)}_{m_j})\;, \eqlabel\new$$
where the parts within a partition satisfy 
$$n^{(j)}_l \geq n^{(j)}_{l+1} + 2j\;.  \eqlabel\difone$$ The different partitions are further subject to the boundary conditions:
$$   n^{(j)}_{m_j} \geq j+ {\rm max}\, (j-i+1,0)+
2j (m_{j+1}+\cdots +  m_{k-1})\;, \eqlabel\condi$$
The length $m$ and the weight $n$ of the partitions enumerated by $f_{k,i}(m,n)$ are related to the above data as follows:
$$n= \sum_{j=1}^{k-1} \sum_{l=1}^{m_j} n^{(j)}_l\qquad {\rm and}\qquad   m= \sum_{j=1}^{k-1} jm_j\;. \eq$$
Clearly, it is because we have a sequence of partitions with a difference condition at distance 1,  i.e., the condition (\difone),  that  the generating function is so easily constructed.

\subsec{A natural generalization}

After deriving this `new' basis of states, we have found that it has actually appeared previously in the literature on vertex operator algebras in  [\ref{G. Georgiev, 
% {\it Combinatorial constructions of modules for infinite-dimensional 
% Lie algebras, I. Principal subspace}, 
J. Pure Appl. Algebra {\bf 112} (1996) 247; 
% (hep-th/9412054) 
  {\it  Combinatorial constructions of modules for infinite-dimensional Lie algebras, II. Parafermionic space}, q-alg/9504024.}\refname\Geo] and in a much more general version.\foot{The basis in [\Geo]  pertains to all models  of the form $\suh(r+1)_k/\uh(1)^r$. For $\suh(2)$, it reduces to the present basis.} Therefore, at the worse, we have provided a conformal-field-theoretical proof of a result already established by means of  vertex-operator-algebra techniques. But we would like to stress the remarkable simplicity of our argument which, by itself, justifies its presentation.

In addition to be simple, our approach seems to have an important potential for generalization. 
This is   illustrated here by the study of the graded parafermions (untreated in [\Geo]), presented in section 4. In that case, two multi-parafermion  bases are derived. One of the resulting generating function is new and it leads to a novel fermionic character formula for graded parafermions.

\newsec{The $\Z_k$ parafermionic models}

The parafermionic conformal algebra is spanned by $k-1$ parafermionic fields $\psi_r$, $r=0,1,\cdots, k-1$, with dimension
$$h_r= r\left(1-{r\over k}\right)\;. \eq$$
Note that $\psi_0=I$, the identity field.
For the present purpose, we will only need the following OPE  [\ref{A.B.
Zamolodchikov and V.A. Fateev, Sov. Phys. JETP {\bf 43} (1985) 215.}\refname\ZFa]:
$$
\psi_r (z) \,\psi_{s} (w) \sim {c_{r,s}\over  (z-w)^{2rs/k}}\;\left[
 \psi_{r+s} (w) + {{r\over r+s}  } (z-w) \d \psi_{r+s}+\cdots\right] \qquad (r+s\leq k) \;,
 \eqlabel\zkope$$
 % \psi_r (z) \,\psi_{s} (w) &\sim { c_{r,s}\over 
% (z-w)^{2(k-r-s)+2rs/k} }\;
  % \psi_{r+s-k} (w)  \qquad (r+s> k) \cr
% \psi_r (z) \,\psi_{k-r}(w) &\sim {1\over (z-w)^{2r(k-r)/k}}
% \left[I+(z-w)^2 {2 h_{\psi_r} \over c}\, T(w) +\cdots\right]\;. 
% \cr} \eqlabel\zkope$$
% The  central charge as well as 
where the structure constants $c_{r,s}$ are fixed by associativity [\ZFa] (their explicit form will not be needed here).
%  $$
 % c={2(k-1)\over (k+2)} \;\;, \qquad
%  c^2_{r,s}={ (r+s)!  (k-r)!  (k-s)! \over 
%  r! s! (k-r-s)! k!}\;.
%  \eqlabel\ctestruc$$ 
 % where the last expression holds for $r+s<k$. 
%  {\bf Verifier le deuxieme $c_{r,s}$. }

Recall that the decomposition of the parafermionic field in modes depends upon the field on which it acts
[\ZFa]. It is essentially fixed by the mutual locality, which is the phase that results from the
substitution
$z\rw ze^{2\pi i}$, denoted  $e^{2\pi i \gamma}$.  The OPE $\psi_r(z)\psi_s(w)$ indicates that the mutual
locality coefficient of $\psi_r$ and $\psi_s$, denoted $\gamma_{r,s}$,  is
$\gamma_{r,s}= -2rs/k$. From the mutual locality coefficient, we can introduce a charge $q$, defined as
$$\gamma_{r,s} = -{q_rq_{s}\over 2k }\;. \eqlabel\chardef$$ The charge is normalized by setting $q_1= 2$, so that $q_r=2r$.  The mutual locality
coefficient of
$\psi_r$ and
$\phi_q$, a generic field of charge
$q$, will then be $-rq/k$.
Therefore, the mode decomposition of $\psi_r$ acting on an arbitrary field $\phi_q$ reads:
$$ \psi_r(z)\phi_{q}(0)  = \sum_{m=-\y}^\y
z^{-rq/k-m-r}A^{(r)}_{r(r+q)/k+m}\, \phi_{q}(0)  \;, \eqlabel\modep$$
the fractional power of $z$ being fixed by the mutual locality.
% A  similar expression holds for the decomposition of
%  $\psi^\dagger_r$: $$
% \psi_r^\dagger (z)\phi_{q}(0)  = \sum_{m=-\y}^\y
% z^{rq/k-m-r} A_{r(r-q)/k+m}^{(r)^\dagger}\,\phi_{q}(0)\eqlabel\modepp$$
% In addition to be compatible with the mutual locality, 
% these expansions also ensure that the
% dimension of a parafermionic mode is the negative of its index.
% Below, we will need the
%  inverted version of (\modep):
 % $$ A^{(r)}_{r(r+q)/k+m} \phi_{q} (0) = {1\over 2 \pi i}\oint_0 dz\,
 %   z^{rq/k+m+r-1}\,\psi_r (z)\, \phi_{q} (0)
 % \cr
%  & A_{r(r-q)/k+m}^{(r)^\dagger} \phi_{q} (0) ={1 \over 2 \pi i} \oint_0 
% {dz }\,  z^{-qr/k+m+r-1}\,\psi^\dagger_r (z)\,
% \phi_{q}(0) \cr}
% \;. \eqlabel\modepin$$ 

 In the following, and in agreement with our previous works [\ref{P. Jacob and P. Mathieu, Nucl. Phys.
{\bf B 587} (2000) 514.}\refname\JMa, \JMb], the fractional part of the modes is
omitted (being fixed unambiguously by the charge of the field or  state on which it acts) and this is
indicated by  calligraphic symbols, i.e.,\foot{This notation simplifies considerably the writing but
it should be kept in mind that the conformal dimension of the mode is no longer given by minus its
index. Note that here $|\phi_{q}\R$ stands for an arbitrary state of charge $q$.}
$$\A^{(r)}_n |\phi_{q}\R\equiv  A^{(r)}_{n+r(r+q)/k}|\phi_{q}\R\; . \eq$$
% \qquad \A_n^{(r)^\dagger}|\phi_{q}\R\equiv 
% A_{n+(1-q)/N}^{(r)^\dagger}|\phi_{q}\R\eq$$

A form of the  commutation relation between the 
$\A^{(r)}$ and $\A^{(s)}$ modes for $r+s\leq k$  follows from the computation of the integral 
$${1\over (2 \pi i)^2}\oint_{C_1} dw\,\oint_{C_2} dz\,
  z^{qr/k+n}\, w^{qs/k+m}\,(z-w)^{-2+2rs/k}\,
\psi_r (z)\, \psi_s (w)\, \phi_{q} (0)\;, \eqlabel\preint$$
by standard contour
deformation\foot{The integral for 
$C_2$ circulating around $w$ while  $C_1$ is a small contour around the origin is compared to 
the difference of two
contours, one with
$|z|>|w|$ and the other with $|z|<|w|$. Note that in the later case, $\psi_r$ passes over $\psi_s$ and
this produces a phase factor $(-1)^{-2rs/k}$ that is partly canceled by the one coming from
$(z-w)^{-2+2rs/k}\rw (-1)^{2rs/k}(w-z)^{-2+2rs/k}$.}.
The result is (omitting the state associated to $\phi_{q}(0)$ on which it acts): 
$$ \sum_{l=0}^{\infty} C^{(l)}_{2rs/k-2} \left[ \A^{(r)}_{n-l-r-1}
\A_{m+l-s+1}^{(s)} - \A_{m-l-s-1}^{(s)} \A^{(r)}_{n+l-r+1}
 \right]   =a\;  c_{r,s}\,\A^{(r+s)}_{n+m-r-s+1}\;, 
\eqlabel\comma$$ 
 where 
 $$ C_t^{(l)}=  {\Gamma(l-t)\over l!\,
\Gamma(-t)}  \;\quad, \qquad a= \left( {ns-mr\over r+s}\right) \;. \eq$$
In the above integral, the power of $z-w$ is chosen in order to pick up precisely
the first two non-vanishing terms of the OPE (in contradistinction with the usual presentation of the commutation relation where only the first non-vanishing term is picked out). We stress that this is made possible by the fact that  in the module $A^{(r+s)}_{-r-s}|0\R$, there is a single descendant of relative charge 0 and relative level 1 and it is proportional to  $L_{-1}A^{(r+s)}_{-r-s}|0\R$. Now the reason for which we pick up these two terms is to extract the maximal amount of constraint from the commutator without generating new types of fields, that is, fields other than $\psi_{r+s}$.\foot{For instance,   the next subleading term of the OPE would involve the new field $(T\psi_{r+s})$.}
% \n {\bf Ajouter la cte de proportionalite `cte' a droite.}

% We will also need  the commutation relation between
% $\A^{(r)}_u$ and $\A^{(k-r)}_v$ 
% for which we  consider  the double integral
% $${1\over (2 \pi i)^2}\oint_{C_1} dw\,\oint_{C_2} dz\,
%   z^{qr/k+n}\, w^{-qr/k+m}\,(z-w)^{2r-2-2r^2/k}\,
% \psi_r (z)\, \psi_r^\dagger (w)\, \phi_{q} (0)\eqlabel\preinttt$$
%  The resulting commutation reads:
% $$\eqalign{
%  \sum_{l=0}^{\infty} &C^{(l)}_{2r-2-2r^2/k} \left[ \A^{(r)}_{n-l+r-2}
% \A_{m+l-r+2}^{(r)^\dagger} + \A_{m-l+r-1}^{(r)\dagger}  \A^{(r)}_{n+l-r+1}
%  \right]  \cr & \quad = {\rm cte'} \; {1 \over 2}
% (n+{qr\over k})(n-1+{qr\over k})
% \delta_{n+m,0}  \cr} \eqlabel\comisa$$
%\n {\bf Fixer la cte}

 Denote  the parafermionic primary fields by $\{\vph_\ell\,| \ell= 0,\cdots ,k-1\}$ [\ZFa,\JMa]. To each
primary field, there corresponds a highest-weight state
$|\vph_\ell\R$. In particular, $|0\R=|\vph_0\R$.  The parafermionic highest-weight  
conditions read
$$\A_{-n-r}^{(r)} | \vph_\ell 
\rangle 
=0  \qquad {\rm for}\quad n<{\rm max}\, (r-k+\ell ,0) \eqlabel\hiwe$$
Note that $ \psi_r(0)| 0 \rangle =\A^{(r)}_{-r}| 0\rangle \propto (\A_{-1})^r 
| 0 \rangle$.

%YYY
% We will also need some information concerning singular vectors [\JMa] . The highest-weight state
% $|\vph_\ell\R$ has the following two primary singular vectors:
% $$|\chi_\ell\R\ = \A_{-1}^{k-\ell+1}|\vph_\ell\R \, ,\qquad
% |\chi'_\ell\R\ = (\B_{0})^{\ell+1}|\vph_\ell\R \eqlabel\sig$$ 
% \n {\bf Comment reexprimer ca en terms des diif. modes? Necessaire ? A voir}

\newsec{A multi-parafermion basis of states}

We look for a basis of states constructed out of the $k-1$ parafermionic modes, 
that is, a  basis of the form
$$\A^{(1)}_{-n^{(1)}_1}\cdots \A^{(1)}_{-n^{(1)}_{m_1}} \A^{(2)}_{-n^{(2)}_1} \cdots\A^{(2)}_{-n^{(2)}_{m_2}}\cdots \A^{(k-1)}_{-n^{(k-1)}_1} \cdots\A^{(k-1)}_{-n^{(k-1)}_{m_{k-1}}} \; |\varphi_\ell\R\;. \eq$$
The goal being to determine the set of independent states for a sequence  of this type, one needs to find those conditions on the  indices $n^{(j)}_l$ that would avoid over counting. These conditions are to be fixed by
the commutation relations. In those relations, 
 we can clearly set to zero those terms already considered. In particular, since each type of modes $\A^{(p)}_{n}$ for $0\leq p\leq k-1$ is considered successively, we can drop their contribution on the rhs of the commutation relations (\comma) with $p=r+s$ and set:
 $$ \sum_{l=0}^{\infty} C^{(l)}_{2rs/k-2} \left[ \A^{(r)}_{n-l-r-1}
\A_{m+l-s+1}^{(s)} - \A_{m-l-s-1}^{(s)} \A^{(r)}_{n+l-r+1}
 \right]   \sim 0\;.
\eqlabel\commA$$ 

Let us now look at the consequences of these simplified relations. Consider first the string of $ \A^{(1)}$ modes and set $r=s=1$ in (\commA):
$$ \sum_{l=0}^{\infty} C^{(l)}_{2/k-1} \left[ \A^{(1)}_{n-l-2}
\A_{m+l}^{(1)} - \A_{m-l-2}^{(1)} \A^{(1)}_{n+l}
 \right]   \sim 0\;. 
\eqlabel\commad$$
This shows that moving a $ \A^{(1)}$ mode to the  right of another $ \A^{(1)}$ mode produces a shift $\Delta$ of its mode index by at least 2, that is,  
$$\Delta= n+l-(n-l-2)=2l+2\geq 2\;. \eq$$ Therefore, the $ \A^{(1)}$ sequence of independent descendants takes the form
$$\A^{(1)}_{-n^{(1)}_1}\cdots \A^{(1)}_{-n^{(1)}_{m_1}} |\cdots\R \qquad {\rm with} \qquad n^{(1)}_{l}\geq n^{(1)}_{l+1}+2 \;. \eq$$
In other words, because we have a shift by at least $2$ in the commutation relation (\commad), we have a difference condition of 2  between adjacent parts. Moreover, the highest-weight condition requires $n_{m_1}\geq 1$.  But this inequality on $ n_{m_1}$ is bounded to be modified by the presence of higher modes. 
Indeed, consider next the commutation of $\A^{(1)}$ and $\A^{(2)}$:
% We can write 
$$ \sum_{l=0}^{\infty} C^{(l)}_{4/k-2} \left[ \A^{(1)}_{n-l-2}
\A_{m+l-1}^{(2)} - \A_{m-l-3}^{(2)} \A^{(1)}_{n+l}
 \right]  \sim 0\;. 
\eqlabel\commadd$$ 
% where the 0 on the rhs is justified by the fact that  the 
% contribution of the different parafermionic modes , will be considered in turn.
We see that by moving a $\A^{(1)}$ mode to the right of a $\A^{(2)}$ mode generates a shift of at least 2. Therefore, when the $\A^{(1)}$'s are preceded by a string of $m_2$ $\A^{(2)}$ modes, the $\A^{(1)}$ indices are shifted by the additional term $2m_2$. More generally,  the relation
$$ \sum_{l=0}^{\infty} C^{(l)}_{2r/k-2} \left[ \A^{(1)}_{n-l-2}
\A_{m+l-r+1}^{(r)} - \A_{m-l-r-1}^{(r)} \A^{(1)}_{n+l}
 \right]  \sim 0\;. 
\eqlabel\commaddd$$ shows that passing $\A^{(1)}$ over $\A^{(r)}$ (for any $r>1)$ generates a shift of at least 2. Therefore, the presence of higher modes to the right of the $\A^{(1)}$ ones induces a shift of all the $\A^{(1)}$ modes by $2(m_2+\cdots +m_{k-1})$. This reproduces (\condi) for $j=1$ up to the $\ell$-dependent boundary term.

Consider now the constraints on the $ \A^{(2)}$ modes. The highest-weight condition requires $n^{(2)}_{m_2}\geq 2$. Now, since we have already taken into account the  commutation of $ \A^{(2)}$ with  $\A^{(1)}$, it suffices to consider that of $ \A^{(2)}$ with $\A^{(r)}$ modes for $r\geq 2$. But actually, the resulting constraints  for those cases cannot be obtained by the commutation relations since the various types of modes have already been generated by the commutators that involve $\A^{(1)}$. To be explicit, we must take due care of the fact that, say $\A^{(1)}\A^{(3)}\sim \A^{(2}\A^{(2)}\sim \A^{(4)}$. Instead, constraints on higher modes have to be determined by the associativity requirement. Since 
$\A^{(2)}\sim \A^{(1)} \A^{(1)}$,  moving $\A^{(r)}$ past a $\A^{(2)}$ mode induces a shift of at least 4 (2 for each $\A^{(1)}$) for any $r\geq 2$. 
Therefore, within the string of $ \A^{(2)}$ modes, we have a difference condition of $4$ between adjacent modes (that follows by considering $r=2$) and a global shift of 4 times the number of other type of modes at its right, that is, $4(m_3+\cdots m_{k-1})$ (from the  $r>2$ cases). This yields (\difone) and (\condi) for $j=2$ (again disregarding the $\ell$-part of the boundary condition).

More generally, to extract the constraints for the commutation of $ \A^{(i)}$ and $ \A^{(j)}$ by associativity,  in order to find the less restrictive conditions, we expand the mode with smallest index ($i$ or $j$) in terms of $\A^{(1)}$ modes. We then find that the resulting shift is  2 min$(i,j)$ for the other mode (with index max $(i,j)$). This readily shows that the parts 
$n^{(j)}_{l}$ satisfy
$$  n^{(j)}_{l}\geq n^{(j)}_{l+1}+2j \;,\eq$$ together with
$$\qquad n^{(j)}_{m_j}\geq j+2j(m_{j+1}+\cdots +m_{k-1})\;.\eq$$

Let us now construct the  generating function for this basis of states, ignoring in the first step the boundary condition on $\ell$. 
Let us first take into account the contribution of the $\A^{(1)}$ modes. It is given by enumerating ordinary partitions of length at most $m_1$, all shifted by  the staircase of weight:
$$\sum_{l=0}^{m_1-1} [2 l+1+2(m_2+\cdots m_{k-1})]= m_1^2+2m_1(m_2+\cdots m_{k-1})\;. \eq$$
By introducing the dummy variable $z_1$ to keep track of the number of $\A^{(1)}$ modes, we have
$$\sum_{m_1\geq 0} z_1^{m_1} {q^{m_1^2+2m_1(m_2+\cdots m_{k-1} )}\over (q)_{m_1}}\eq$$
More generally, the contribution of the $\A^{(j)}$ modes is obtained by enumerating ordinary partitions of length at most $m_j$ shifted by  the staircase of step $2j$, whose weight, properly modified by the presence of the number of modes of higher type (i.e., $r>j)$, is
$$ j \sum_{l=0}^{m_j-1} [2l+1+2(m_{j+1}+\cdots m_{k-1})]= jm_j^2+2jm_j(m_{j+1}+\cdots m_{k-1})\;.\eq$$
This contributes to the factor
$$\sum_{m_j\geq 0} z_j^{m_j} {q^{jm_j^2+2jm_j(m_{j+1}+\cdots m_{k-1} )}\over (q)_{m_j}}\;. \eq$$
Summing up all terms, we end up with the following generating function
$$\sum_{m_1,\cdots,m_{k-1}=0}^\y {
q^{N_1^2+\cdots+ N_{k-1}^2}\; \prod_{j=1}^{k-1}z_j^{m_j} \over (q)_{m_1}\cdots (q)_{m_{k-1}}
}\;,\eqlabel\defFaa$$
where the $N_j$ are defined in (\efNL). We can introduce a single variable to keep track of the relative charge of the descendant states instead of the length of its various parts by defining
$z_j= z^j$. This leads to
$$\sum_{m_1,\cdots,m_{k-1}=0}^\y {
q^{N_1^2+\cdots+ N_{k-1}^2}\; z^{N_1+\cdots N_{k-1}}\over (q)_{m_1}\cdots (q)_{m_{k-1}}
}\;. \eqlabel\defFa$$

Let us now take care of the boundary condition that characterizes the different modules. This is a further constraint that ensures that the first $r$-type descendant of a highest-weight state labeled by $\ell$, namely  $\A^{(r)}_{-n^{(r)}_{m_r} }|\varphi_\ell\R$, does not have a negative dimension. This is prevented by requiring that (cf. (\hiwe))
 $$n^{(r)}_{m_r} \geq r+ {\rm max}\, (r-k+\ell ,0)\; .\eqlabel\bdry$$
% These conditions are  analogous to the singular vector 
% constraint $\left[\A_{-1}^{(1)}\right]^{k-\ell+1}|\varphi_\ell\R=0$.
The bound (\bdry)  produces a  global shift for all the  indices of type $r$ such that $r-k+\ell>0$. Summing their contribution generates the weight factor $$\sum_{r=1}^{k-1} {\rm max}\, (r-k+\ell,0) \, m_r= m_{k-\ell+1}+2m_{k-\ell+2}+\cdots ({k-1})m_{k-1}=N_{k-\ell+1}+\cdots N_{k-1}\;. \eq$$
This reproduces precisely the linear term in the  exponent of $q$ in (\defF) for $i=k-\ell+1$. We have thus recovered the function $F_{k,i}(z;q)=F_{k,k-\ell+1}(z;q)$.

Note finally that by reinserting the fractional contribution of the modes (e.g., as described in section 5.2 of [\JMb]) one recovers the Lepowsky-Primc character formula for the $\Z_k$ parafermionic models [\LP].

\newsec{New quasi-particle bases for graded parafermions}

\subsec{Preliminary remarks on graded parafermions}

Graded parafermions [\ref{J. M. Camino, A. V. Ramallo and
J. M. Sanchez de Santos, Nucl.Phys. {\bf B530} (1998) 715.}\refname\CRS] are associated  to the coset $\osp(1,2)_k/\uh(1)$. The corresponding chiral algebra is generated by $2k-1$ parafermions  ${\tilde\psi}_r,\; r=0,\frac12,1,\cdots, k-\frac12$, of dimension
$${\tilde h}_{r}= r \left( 1-{ r \over k}\right) +{\e_r\over 2}\;, \eq$$
where $\e_r=0$ if $r$ is integer and 1 otherwise.  The conformal dimension of the lowest dimensional
parafermion
${\tilde\psi}_{1/2}$ is thus $1-1/4k$.  The defining OPE  reads $(r+s\leq k)$
$$
{\tilde\psi}_r (z) \,{\tilde\psi}_{s} (w) \sim {{\tilde c}_{r,s}\over  (z-w)^{2rs/k+\e_r\e_s}}\;
[{\tilde\psi}_{r+s} (w)+\cdots] \;. \eqlabel\zkope$$
Notice that for $r+s$ half-integer, there are more than one descendant-field at level 1.
The mode decomposition is defined as
$${\tilde \psi}_r(z)\phi_{q}(0)  = \sum_{m=-\y}^\y
z^{-rq/k-m-r-\e_r/2}{\tilde A}^{(r)}_{r(r+q)/k+m}\, \phi_{q}(0)  \;, \eqlabel\modep$$
As before, we will avoid writing the fractional part of the modes explicitly. 
The primary fields ${ \tilde \varphi}_\ell$ are parametrized by an integer $\ell$ such that $0\leq \ell\leq k$.  The highest-weight conditions (that ensure the absence of any negative-dimensional descendants) are:
$${\tilde \A}^{(r)}_{-r-\e_r/2-n} |{ \tilde \varphi}_\ell\R=0\qquad{\rm if }\qquad n< {\rm max}\, \left(r-{\e_r\over 2}-k+\ell,0\right)\eqlabel\BDry$$

\subsec{A first graded multi-parafermion basis}

The first basis we look for is of the form
$${\tilde\A}^{(1/2)}_{-n^{(0)}_1}\cdots {\tilde\A}^{(1/2)}_{-n^{(0)}_{m_{0}}}{\tilde\A}^{(1)}_{-n^{(1)}_1}\cdots {\tilde\A}^{(1)}_{-n^{(1)}_{m_1}} {\tilde\A}^{(2)}_{-n^{(2)}_1} \cdots{\tilde\A}^{(2)}_{-n^{(2)}_{m_2}}\cdots {\tilde\A}^{(k-1)}_{-n^{(k-1)}_1} \cdots{\tilde\A}^{(k-1)}_{-n^{(k-1)}_{m_{k-1}}}\; |\tilde{\varphi}_\ell\R \;. \eq$$
We then have to find the constraints on the different type of indices by considering the commutation relations.  Consider first the commutator between two ${\tilde\A}^{(1/2)}$  modes. For this, since the basis includes the ${\tilde\A}^{(1)}$  modes and because the ${\tilde\A}^{(1)}$ module has a single zero-charge descendant at level 1, we can pick up the first two non-vanishing terms in the OPE. This results into [\CRS, 
\ref{P. Jacob and P. Mathieu,  Nucl. Phys. 
{\bf B630} (2002) 433.}\refname\JMc]:
$$\sum_{l\geq0} C_{1/2k-1}^{(l)}  [ {\tilde \A}^{(1/2)}_{n-l-1} {\tilde \A}^{(1/2)}_{m+l}   - {\tilde \A}^{(1/2)}_{m-l-1} {\tilde \A}^{(1/2)}_{n+l} ] \sim 0\; .\eq $$
This indicates a difference 1 between adjacent modes ${\tilde \A}^{(1/2)}$:
$$n^{(0)}_l\geq n^{(0)}_{l+1} +1\;.\eq$$
The condition (\BDry) yields $n^{(0)}_{m_0}\geq 1$.  Next, we consider the commutator of ${\tilde\A}^{(1/2)}$  with ${\tilde\A}^{(r)}$  for $r$ integer. Since we do not take into account the modes ${\tilde\A}^{(r+1/2)}$ in this basis, we must avoid picking up any non-vanishing terms on the rhs of the corresponding OPE. The strongest constraint we get with this restriction is
$$\sum_{l\geq0} C_{r/k+1}^{(l)}  [ {\tilde \A}^{(1/2)}_{n-l+1} {\tilde \A}^{(r)}_{m+l-r+1}   - {\tilde \A}^{(r)}_{m-l-r+1} {\tilde \A}^{(1/2)}_{n+l+1} ] = 0\eq $$
This implies that the smallest mode-shifting we can get when a $ {\tilde \A}^{(1/2)}$ mode is moved past a  $ {\tilde \A}^{(r)}$ mode is zero. That indicates that the presence of higher modes at the right of the $  {\tilde \A}^{(1/2)}$ string does not affect the latter modes, that is, it does not modify the bound $n^{(0)}_{m_0}\geq 1$. As a result, there will be  no interacting term of the type $m_0m_r$ in the generating function.

 For the  other modes, the analysis is similar to the one pertaining to  the non-graded case. 
Hence, (\difone) and (\condi) still hold for $j\geq 1$.
 
 This basis has the following generating function:
 $$\sum_{m_{0}, m_1,\cdots,m_{k-1}=0}^\y {
q^{m_{0}(m_{0}+1)/2+N_1^2+\cdots+ N_{k-1}^2+N_{k-\ell+1}+\cdots N_{k-1}}\; z_{0}^{m_{0}}\prod_{j=1}^{k-1}z_j^{m_j}  \over  (q)_{m_{0}}(q)_{m_1}\cdots (q)_{m_{k-1}} }\;.\eqlabel\defFab$$
Setting  $z_0= z$, $z_j= z^{2j}$ and summing over the $m_{0}$ modes, this becomes:
$$(-zq)_\y \sum_{ m_1,\cdots,m_{k-1}=0}^\y  {
q^{N_1^2+\cdots+ N_{k-1}^2+N_{k-\ell+1}+\cdots N_{k-1}} z^{2(N_1+\cdots N_{k-1})}\over  (q)_{m_1}\cdots (q)_{m_{k-1}} }= (-zq)_\y F_{k,k-\ell+1}(z^2;q)\;,\eqlabel\defFac$$
which is precisely the result obtained in  [\ref{L. B\'egin, J.-F. Fortin, P. Jacob and P. Mathieu, Nucl. Phys {\bf B659} (2003) 365.}\refname\BFJM] (cf. eq. (3.24)). The simplicity of this derivation contrasts heavily with that in the later reference, which requires the enumeration of restricted jagged partitions [\JMc, \ref{
J.-F. Fortin, P. Jacob and P. Mathieu, {\it Jagged partitions}, Ramanujan J., to appear; math.CO/0310079.}].

\subsec{A second graded mutli-parafermion basis}

The second basis we consider involves all graded parafermionic modes, that is,
$${\tilde\A}^{(1/2)}_{-n^{(1/2)}_1}\cdots {\tilde\A}^{(1/2)}_{-n^{(1/2)}_{m_{1/2}}}{\tilde\A}^{(1)}_{-n^{(1)}_1}\cdots {\tilde\A}^{(1)}_{-n^{(1)}_{m_1}} {\tilde\A}^{(3/2)}_{-n^{(3/2)}_1} \cdots{\tilde\A}^{(3/2)}_{-n^{(3/2)}_{m_{3/2} }}\cdots {\tilde\A}^{(k-1/2)}_{-n^{(k-1/2)}_1} \cdots{\tilde\A}^{(k-1/2)}_{-n^{(k-1/2)}_{m_{k-1/2}}}\; |\tilde{\varphi}_\ell\R \;. \eq$$
Let us note readily the boundary condition on each index $n^{(j)}_{m_j}$ that results from (\BDry):
$$ n^{(j)}_{m_j} \geq j+{\e_j\over 2}+{\rm max}\, \left(j-{\e_j\over 2}-k+\ell,0\right)\;. \eq$$

Again we start by considering the commutation relation between   the $ {\tilde \A}^{(1/2)}$ and ${\tilde \A}^{(r)}$ modes, where now $r$ can be  both integer and half-integer.  For $r=1/2$, the analysis of the previous subsection still holds. Thus, here again, the $ {\tilde \A}^{(1/2)}$ modes have to be distinct. 
For $r> 1/2$, if the produced module  ${\tilde \A}^{(r+1/2)}$ has a single zero-relative-charge descendant at level 1, we can pick up two non-vanishing terms on the rhs of the OPE. This is the case when $r$ is half-integer. 
The relevant commutation relations is then
$$\sum_{l\geq0} C_{r/k-1}^{(l)}  [ {\tilde \A}^{(1/2)}_{n-l-1} {\tilde \A}^{(r)}_{m+l-r}   - {\tilde \A}^{(r)}_{m-l-r-1} {\tilde \A}^{(1/2)}_{n+l} ] \sim 0\;. \eq$$
This implies a shift of 1 in $ {\tilde \A}^{(1/2)}$ modes for each $ {\tilde \A}^{(r)}$ modes at its right, with $r$ half-integer. For $r$ integer, it turns out that there are generically three zero-relative-charge descendant at level 1 in the module  ${\tilde \A}^{(r+1/2)}$.\foot{To see this neatly, take $k$ large. The character (normalized such that the leading term is 1) of the vacuum module of relative charge $2r$ (that is, the module of $\A_{-r}^{(r)}|0\R$) is given by (for $r>1/2)$:
$$\chi_{2r}(q)\approx V_{2r}(q)-V_{2r+1}(q)= 1+(1+2\e_r) q+\cdots$$
where $V_t$ denotes the Verma module of relative charge $t$ (cf. eqs (5.4)-(5.6) and (5.12)-(5.13) of [\BFJM]).}
Therefore, only the first non-vanishing term must be considered in the OPE $\psi_{1/2}(z)\psi_r(w)$. This gives
$$\sum_{l\geq0} C_{r/k-1}^{(l)}  [ {\tilde \A}^{(1/2)}_{n-l-1} {\tilde \A}^{(r)}_{m+l-r+1}   - {\tilde \A}^{(r)}_{m-l+r} {\tilde \A}^{(1/2)}_{n+l} ] \sim 0\;, \eq$$
and again this implies a shift of 1 in $ {\tilde \A}^{(1/2)}$ modes for each $ {\tilde \A}^{(r)}$ modes at its right, with $r$ integer. Associativity (decomposition of higher modes into a product of $ {\tilde \A}^{(1/2)}$  ones) show that when $ {\tilde \A}^{(r)}$ is passed over a  ${\tilde \A}^{(s)}$, there is a difference 2 min $(r,s)$. 
This is thus a difference of $2r$  between the  $ {\tilde \A}^{(r)}$ modes and a shift of $2r$ for each higher modes at its right. When summing over the contribution of the $r$ modes, this generates the weight 
$$rm_r^2 +2r m_r (m_{r+1/2}+m_{r+1}+\cdots m_{k-1/2})\eq$$

The $\ell$-dependent boundary term that has been ignored so far is evaluated as in the non-graded case:
$$\sum_{r=1}^{k-1/2} {\rm max}\, \left(r-{\e_r\over 2} -k+\ell,0\right) \, m_r= (m_{k-\ell+1}+m_{k-\ell+3/2} )+\cdots  +({\ell-1})(m_{k-1}+ m_{k-1/2})\equiv {\tilde L}_{k-\ell+1}\;. \eqlabel\defL$$

The resulting generating function is thus:
$$\sum_{m_{1/2}, m_1,m_{3/2}\cdots,m_{k-1/2}=0}^\y {
q^{\frac12({\tilde N}_{1/2}^2+{\tilde N}_1^2+\cdots+ {\tilde N}_{k-1/2}^2 +M_{1/2})+{\tilde L}_{k-\ell+1}}\; \prod_{j=1/2}^{k-1/2}z_j^{m_j}  \over  (q)_{m_{1/2}}(q)_{m_1}\cdots (q)_{m_{k-1/2}} }\;.\eqlabel\defFabb$$
where $${\tilde N}_{j}= m_j+m_{j+1/2}+\cdots + m_{k-1/2}= M_j+M_{j+1/2}\;, \eq$$
with
$$M_j= m_j+m_{j+1}+\cdots +m_{k-1+\e_j/2}\;, \eq$$
and ${\tilde L}_{k-\ell+1} $ defined in (\defL).
% $${\tilde L}_{k-\ell+1}= M_{k-\ell+1}+\cdots +M_{k-1}
% +M_{k-\ell+3/2}+\cdots +M_{k-1/2}\eq$$
With $z_j=z^{2j}$, the $z$ factor reduces to $z^{{\tilde N}}$ where ${\tilde N}= \sum 2jm_j$ and we have
$$G_{k,k-\ell+1}(z;q)= \sum_{m_{1/2}, m_1,m_{3/2}\cdots,m_{k-1/2}=0}^\y {
q^{\frac12({\tilde N}_{1/2}^2+{\tilde N}_1^2+\cdots+ {\tilde N}_{k-1/2}^2 +M_{1/2})+{\tilde L}_{k-\ell+1}}\; z^{{\tilde N}} \over  (q)_{m_{1/2}}(q)_{m_1}\cdots (q)_{m_{k-1/2}} }\;.\eqlabel\defFabb$$

\subsec{A  generalized  Rogers-Ramanujan identity}

% We stress that $G_{k,k-\ell+1}(z;q)$  
% seems to be a new generating function.  
The equivalence of the two new graded bases implies the equality (with $i=k-\ell+1$):
$$G_{k,i}(z;q)= (-zq)_\y F_{k,i}(z^2;q)\; . \eqlabel\newRRf$$
For $z=1$, the rhs has a product form (cf. [\ref{
J.-F. Fortin, P. Jacob and P. Mathieu, Electronic J. Comb.
  {\bf 12} (2005) R12.}] Theorem 11).
This and the above equality lead  to the following generalization of the Rogers-Ramanujan identity:\foot{Multiple sums similar but not identical to $G_{k,i}(z;q)$ have been conjectured in [\ref{E. Melzer, {\it Supersymmetric Analogs of the Gordon-Andrews Identities, and Related TBA Systems}, hep-th/9412154.}\refname\Mel] as fermionic expressions for the Ramond characters of the  superconformal minimal model ${\cal SM}(2,4k)$. (These identities have been subsequently proved in [\ref{D. Bressoud, M. Ismail and D. Stanton, Ramanujan J. {\bf 4} (2000) 435.}] -- see also Theorem 4.4 of [\ref{S.O. Warnaar, Discrete Math. {\bf 272} (2003) 215.}\refname\Ole]). Note that if we relabel our $m_j$ as $m_{2j}$, and set $m_{2(k-\ell+1)}= m_s$ (so that $s$ is even) together with $m_{2k-1}=0$ in our formula, we recover the expression given in the second line of eq (2.6) in [\Mel]. This signals an unexpected relation between the ${\cal SM}(2,4k)$ models and the $\Z_k$ graded parafermions. The present analysis, in the light of the recent work [\ref{J.-F. Fortin, P. Jacob and P. Mathieu, J. Phys. A: Math. Gen. {\bf 38} (2005) 1699.}], provides a possible path for an alternative proof of these  identities.}
$$\sum_{m_{1/2}, \cdots,m_{k-1/2}=0}^\y {
q^{\frac12({\tilde N}_{1/2}^2+{\tilde N}_1^2+\cdots+ {\tilde N}_{k-1/2}^2 +M_{1/2})+{\tilde L}_{i}}\;  \over  (q)_{m_{1/2}}(q)_{m_1}\cdots (q)_{m_{k-1/2}} }=  \prod_{n=1}^\infty  (1+ q^n)  \prod_{n\not
= 0,
\pm i
\;{\rm mod}\, (2k +1)}^\infty  (1- q^n)^{-1}  \;. \eqlabel\defbB$$
% with $${\tilde L}_{k-\ell+1}= M_{k-\ell+1}+\cdots +
% M_{k-1}+M_{k-\ell+3/2}+\cdots +M_{k-1/2}\eq$$
We stress that with the expression we had  previously [\BFJM] for the specialized multi-sum, i.e., $(-q)_\y F_{k,i}(1;q)$,  the factor 
$(-q)_\y =  \prod_{n=1}^\infty  (1+ q^n)$ would cancel on both sides of the `sum=product'  equality 
$$(-q)_\y F_{k,i}(1;q)=  \prod_{n=1}^\infty  (1+ q^n)  \prod_{n\not
= 0,
\pm i
\;{\rm mod}\, (2k +1)}^\infty  (1- q^n)^{-1} \;, \eqlabel\deftri$$  reducing then to  the usual Andrews-Gordon identity. But there is no such cancelation  with (\defbB) (except for the trivial case $k=1$). In particular, for $k=2$, it reads:
$$\sum_{n,m,p=0}^\y {
q^{\frac12n^2+m^2+\frac32p^2+n(m+p)+2mp+(2-i)(m+p)+\frac12(n+p)}\;  \over  (q)_{n}(q)_{m} (q)_{p} }=  \prod_{n=1}^\infty  (1+ q^n) \prod_{n\not=0,\pm i\; {\rm mod}\; 5}{1\over 1-q^n}\;,\qquad\qquad (i=1,2)\
\eqlabel\defbBRR$$
% \over (1- q^{5n+3-i})(1- q^{5n-3+i})]}  \;. \eqlabel\defbBRR$$
% $$\sum_{m_{1/2}, m_1, m_{3/2}=0}^\y {
% q^{\frac12m_{1/2}^2+m_1^2+\frac32m_{3/2}^2+m_{1/2}
% (m_1+m_{3/2})+2m_1m_{3/2}+(2-i)(m_1+m_{3/2})+
%  \frac12(m_{1/2}+m_{3/2})}\;  \over  (q)_{m_{1/2}}
% (q)_{m_1} (q)_{m_{3/2}} }=  \prod_{n=1}^\infty { (1+ q^n) 
% \over (1- q^{5n+3-i})(1- q^{5n-3+i})]}  \;. %\eqlabel\defbBRR$$
Because it involves the modulus 5 on the rhs, this identity  could be viewed as the  fermionic deformation of the original  Rogers-Ramanujan identity (\RRs).

There is a stricking similarity between (\defbB) for $i=k$ and the identity of  Theorem 4.5 of [\Ole]. In fact, Warnaar [{\ref{S.O. Warnaar, private communication.}\refname\Opc] has  shown that these two relations are  essentially equivalent. The sketch of the proof -- which is an analytic counterpart of our conformal-field-theoretical proof of (\newRRf) -- is reported in  Appendix A. 

\subsec{The $\Z_k$ graded multi-parafermion bases: combinatorial formulation}

Taken together, the results of  [\JMc] and the present ones have the following combinatorial interpretation. There is an equality  between the number of partitions described by the following three sets. 

\n 1- The first set corresponds to the jagged partitions $(n_1,\cdots , n_m)$ defined as 
$$
 n_j\geq n_{j+1}-1\;,\qquad  \qquad  n_j\geq n_{j+2}\;, \qquad\qquad   n_m\geq 1\;,\eq
$$ with at most $i=1$ pairs of $01$ 
 and further subject to the following  the $k$-restrictions: 
$$
n_j \geq  n_{j+2k-1} +1 \qquad{\rm or} \qquad
 n_j = n_{j+1}-1 =  n_{j+2k-2}+1= n_{j+2k-1} \;,\eq
$$
for all values of $j\leq m-2k+1$, with $k>1$. 

\n 2- The second set corresponds to a sequence of $k$ ordered partitions $(n^{(0)}, n^{(1)}, n^{(2)}, \cdots ,n^{(k-1)})$ of 
respective lengths $m_0, \,  m_1,\cdots , m_{k-1}$, with 
$$n^{(0)}_l\geq n^{(0)}_{l+1} +1\;,\qquad 
n^{(j)}_l \geq n^{(j)}_{l+1} + 2j\;, \eqlabel\difonee$$ with the different partitions being  further subject to the boundary conditions:
$$  n^{(0)}_{m_0} \geq 1\; ,\qquad   n^{(j)}_{m_j} \geq j+ {\rm max}\, (j-i+1,0)+
2j (m_{j+1}+\cdots +  m_{k-1})\;, \eqlabel\condii$$
with $j\geq 1$.

\n 3- Finally, the third set corresponds to a sequence of $2k-1$ ordered partitions $(n^{(1/2)}, n^{(1)}, n^{(3/2)}, \cdots ,n^{(k-1/2)})$ of 
respective lengths $m_{1/2}, \,  m_1,\cdots , m_{k-1/2}$, with 
$$
n^{(j)}_l \geq n^{(j)}_{l+1} + 2j\; ,  \eqlabel\difonee$$ and the boundary conditions:
$$    n^{(j)}_{m_j} \geq j+ {\e_j\over 2} {\rm max}\, \left(j- {\e_j\over 2} + i+1,0\right)+
2j (m_{j+1/2}+m_{j+1}+\cdots +  m_{k-1/2})\;. \eqlabel\condii$$

%%%%%%%%%

\newsec{Conclusion}

In this work we have  displayed a multi-parafermion basis of states for the $\Z_k$ parafermionic models. The basis elements are in one-to-one correspondence with the set of $k-1$ ordered partitions described in  eqs (\new), (\difone) and (\condi). This is an alternative to the usual description of the basis in terms of partitions restricted by (\usual) [\LP, \JMb].  In the parafermionic context, the argued  equivalence of the two bases leads us to the conclusion that the two sets of partitions, namely (\new)-(\condi) and (\usual), are equinumerous. Clearly, finding a direct bijection would allow us to strip off this elementary derivation of $F_{k,i}$ from any parafermionic dressing.
Moreover, such a bijection might point toward natural `higher-rank' generalizations of the Andrews-Gordon identitity.

As previously pointed out, the `new'  $\Z_k$ basis has already been derived in [\Geo]. We have thus emphasized here the novelty (and simplicity) of the conformal-field-theoretical derivation. As an original extension, two new bases of states for graded parafermions have been displayed. 
Each one leads to a distinct fermionic form of the graded-parafermion characters once the contribution of the fractional part of the parafermionic modes is reinserted. The expression linked to the  basis  involving all parafermionic modes is new.  It is interesting to see that for this  basis, an unusual aspect of the representation theory of the graded parafermions (when compared to the standard $\Z_k$ representation theory) plays a crucial role, which is that some graded parafermionic modules have more than one level-1 descendant of relative-charge zero.

%One of it leads to a new Rogers-Ramanujan-type identity. 

This work offers another illustration of the non-uniqueness of  the  fermionic characters of the irreducible modules in a given model. Here, this is rooted in the non-uniqueness of the quasi-particle basis. There are indeed different choices for the spanning set of creation operators that are compatible with a description of the basis in terms of restriction rules akin to exclusion relations. For the $\Z_k$ models, there are two choices: $\{\A^{(1)} \} $  and $\{ \A^{(1)},\cdots,  \A^{(k-1)} \}$. For the graded case, there are three such sets: $\{ {\tilde \A}^{(1/2)} \} $,  $\{ {\tilde \A}^{(1/2)},{\tilde \A}^{(1)},{\tilde \A}^{(2)},\cdots,  {\tilde \A}^{(k-1)} \}$ and $\{ {\tilde \A}^{(1/2)},{\tilde \A}^{(1)},{\tilde \A}^{(3/2)},\cdots,  {\tilde \A}^{(k-1/2)} \}$. These sets are not necessarily exhaustive since, for instance, one could possibly consider a choice where some parafermionic fields are ignored,\foot{On the analytic side, the argument of Appendix A shows clearly how to eliminate,  from the generating function, an arbitrary set of modes associated to the graded parafermions $\psi_r$ with $r$ half-integer.} or even one involving a mixtures of selected parafermionic modes augmented by the addition the Virasoro  or  higher integer-spin field modes. 

% Finally, we emphasis that the various 
%bases presented here are derived at a foundamental level

 \appendix{A}{The analytic proof of (\newRRf) }

 The equality (\newRRf), that  has been established here by a field-theoretical argument, can also be demonstrated by analytical methods  [\Opc].  The general argument would proceed by a simple extension of Lemma A.1 of  [\Ole]. We will content ourself with the consideration of 
 the $k=2$  case  and indicate at the end how the analysis can be  generalized to $k>2$.  Let us first replace  $m_j$ by $m_{2j}$:
$$G_{2,i}(z;q)= \sum_{m_1,m_2,m_3=0 }^\infty{ q^{\frac12(m_1+m_2+m_3)^2+\frac12(m_2+m_3)^2+\frac12m_3^2+\frac12(m_1+m_3)+(2-i)(m_2+m_3)}  z^{ m_1+2m_2+3m_3} \over 
(q)_{m_1}(q)_{m_2}(q)_{m_3} }\;.\eq$$
Next, we replace $m_2$ by $m_2-m_3$ and use the convention that $1/(q)_n=0$ if  $n<0$
   $$\eqalign{
   G_{2,i}(z;q)&= 
\sum_{m_1,m_2,m_3=0 }^\infty {q^{\frac12(m_1+m_2)^2+\frac12m_2^2+\frac12m_3^2+\frac12(m_1+m_3)+ (2-i)m_2}z^{m_1+2m_2+m_3}\over
(q)_{m_1}(q)_{m_2-m_3}(q)_{m_3} }\cr  
% = \sum_{m_1,m_2,m_3=0 }^\infty {q^{\frac12m_1(m_1+2m_2+1)
% +m_2^2+\frac12(m_3(m_3+1) + (2-i)m_2} z^{m_1+2m_2+m_3}\over 
% (q)_{m_1}(q)_{m_2-m_3}(q)_{m_3}}\cr &
& =\sum_{m_1,m_2=0 }^\infty {q^{\frac12m_1(m_1+2m_2+1)+m_2^2+(2-i)m_2} z^{m_1+2m_2} \over 
(q)_{m_1}(q)_{m_2}} \sum_{m_3=0}^{m_2} {(q)_{m_2}\over 
(q)_{m_2-m_3}(q)_{m_3}} q^{\frac12m_3(m_3+1) } z^{m_3}\;.\cr}\eq
$$
 Using the  $q$-binomial theorem ([\Andr], eq (3.3.6))
 $$\sum_{j=0}^n {(q)_n\over (q)_j (q)_{n-j}} q^{\frac12j(j+1)} x^j = (-xq)_n \;,\eq$$
we can perform the summation over $m_3$ and get
 $$G_{2,i}(z;q)=
\sum_{m_1,m_2=0}^\infty{q^{\frac12m_1(m_1+2m_2+1)+m_2^2+(2-i)m_2}z^{m_1+2m_2}\over 
(q)_{m_1}(q)_{m_2}} (-zq)_{m_2}\;.\eq
$$
We next make use of the Euler relation ([\Andr], eq (2.2.6)):
 $$\sum_{n=0 }^\infty{q^{\frac12n(n-1)} x^n\over (q)_n} = (-x)_\infty\;, \eq$$
 to sum over  $m_1$ with $x=zq^{m_2+1}$:
  $$
  \eqalign{
  G_{2,i}(z;q)
% \sum_{m_1,m_2=0 }^\infty{q^{\frac12m_1(m_1+2m_2+1)+
% m_2^2+(2-i)m_2} z^{m_1+2m_2}\over 
% (q)_{m_1}(q)_{m_2}} (-zq)_{m_2}  
= & \sum_{m_2=0 }^\infty (-zq^{m_2+1})_\infty  (-zq)_{m_2} {q^{m_2^2+(2-i)m_2} z^{2m_2}\over (q)_{m_2}}\cr
 = &(-zq)_{\infty} \sum_{m_2=0 }^\infty{ q^{m_2^2+(2-i)m_2} z^{2m_2}\over (q)_{m_2}}= (-zq)_{\infty} F_{2,i}(z^2;q)\;. \cr} \eq$$
 The generalization  to $k>2$ is straightforward. The odd modes $m_{2j+1}$ for $j>1$ are summed successively, starting from the largest one, by the $q$-binomial theorem, while the sum over $m_1$ is done by the Euler relation. The identity of Theorem 4.5 in [\OleÊ]Ê is similarly related to the  multiple-sum of Andrews. With the suitable addition of a linear term, the latter is thus essentially equivalent to our (\newRRf).  
 
\vskip0.3cm
\centerline{\bf Acknowledgment}

This work was supported by EPSRC (PJ) and NSERC (PM).  We would like to thank O. Warnaar for drawing to our attention  some of the quoted   references from the mathematical literature and, most importantly, for communicating to us the argument of Appendix A. 

\vskip0.3cm

%\vfill\eject
\centerline{\bf REFERENCES}
%\vskip 0.5cm
\immediate\closeout\refs \vskip 0.5cm
   \message{References}\input references
%\vfill\eject

\end

 Omitting the details of the analysis, we can show that the commutation relations [\CRS, 
\ref{P. Jacob and P. Mathieu,  Nucl. Phys. 
{\bf B630} (2002) 433.}\refname\JMc] lead to the  following basis of states:
 $${\tilde\A}^{(1/2)}_{-n^{(0)}_1}\cdots {\tilde\A}^{(1/2)}_{-n^{(0)}_{m_{0}}}{\tilde\A}^{(1)}_{-n^{(1)}_1}\cdots {\tilde\A}^{(1)}_{-n^{(1)}_{m_1}} {\tilde\A}^{(2)}_{-n^{(2)}_1} \cdots{\tilde\A}^{(2)}_{-n^{(2)}_{m_2}}\cdots {\tilde\A}^{(k-1)}_{-n^{(k-1)}_1} \cdots{\tilde\A}^{(k-1)}_{-n^{(k-1)}_{m_{k-1}}}\; |\tilde{\varphi}_\ell\R \;. \eq$$
where 
$$n^{(0)}_l\geq n^{(0)}_{l+1} +1\;, \eq$$
so that the parts of the ${\tilde\A}^{(1/2)}$ string are distinct, while the parts  for $j\geq 1$ still satisfy (\difone). In addition, (\condi) still holds but for $j\geq 1$, while for ${\tilde\A}^{(1/2)}$modes,  there are no boundary condition except $n^{(0)}_{m_{0}}\geq 1$. That the presence of the  higher modes does not induce a shift in the ${\tilde\A}^{(1/2)}$ modes is due to the fact that by commuting a  ${\tilde\A}^{(1/2)}$ mode past  a ${\tilde\A}^{(r)}$ one for $r$ integer does not induce any mode shift in those forms of the commutation relations that do not involve (since they are not taken into account in our basis) the new modes  ${\tilde\A}^{(r+1/2)}$.

&&&&&&&&&&

$$\A^{(1)}_{-n_1}\cdots \A^{(1)}_{-n_\ell} \A^{(2)}_{-n'_1} \cdots\A^{(2)}_{-n'_p}\eq$$
avec 
$$n_i\geq n_{i+1} +2 \qquad n_\ell\geq 1+2p\eq$$ et 
$$ n'_i\geq n'_{i+1}+4\qquad n'_p\geq 2\eq$$
On veut compter d'abord la contribution des modes $\A^{(1)}$, dont le nombre est donne par la puissance de $z_1$. L'escalier contribue pour une puissance de $q$ egale a
$$\sum_{j=0}^{\ell-1} (2j+1 + 2p) = \ell^2+2\ell p\eq$$
La contribution des modes $\A^{(1)}$est donc
$$\sum_{\ell\geq 0} z_1^\ell \;  {q^{\ell^2+2\ell p}\over (q)_\ell}\eq$$
En effet $1/(q)_\ell$ est le nombre de partition de longueur au plus $\ell$, ces partitions sont ajoutees a l'escalier (qui est l'etat fondamental).
Pour la contribution des modes $\A^{(2)}$ dont le nombre est la puissance de $z_2$, on doit sommer l'escalier
$$\sum_{j=0}^{p-1} (4j+2)  = 2p^2\eq$$
On a donc, pour les modes $\A^{(2)}$, le resultat suivant :
$$\sum_{p\geq 0} z_2^p \; {q^{2p^2}\over (q)_p}\eq$$
Au total, on a
$$\sum_{\ell,p\geq 0} z_1^{\ell}z_2^{ p}\;   {q^{\ell^2+2\ell p+2p^2}\over (q)_\ell (q)_p}\eq$$
Si on veut que la puissance de $z$ code la charge, on pose
$z_i= z^i$ et on retrouve
$$\sum_{\ell,p\geq 0} z^{\ell +2p}\;   {q^{\ell^2+2\ell p+2p^2}\over (q)_\ell (q)_p}=
\sum_{m_1,m_2\geq 0} z^{m_1+2m_2}\;   {q^{m_1^2+2m_2^2+2m_1m_2}\over (q)_{m_1}(q)_{m_2} }\eq$$
ce qui est la formule d'Andrews pour le cas $k=3$ (et sans restriction sur le nombre de 1 - soit sans le facteur $L_i$).

Cas general: la base serait de la forme
$$\A^{(1)}_{-n^{(1)}_1}\cdots \A^{(1)}_{-n^{(1)}_{m_1}} \A^{(2)}_{-n^{(2)}_1} \cdots\A^{(2)}_{-n^{(2)}_{m_2}}\cdots \A^{(k-1)}_{-n^{(k-1)}_1} \cdots\A^{(k-1)}_{-n^{(k-1)}_{m_{k-1}}} \eq$$
avec
$$n^{(j)}_i \geq n^{(j)}_{i+1} + 2j \qquad{\rm et }\qquad   n^{(j)}_{m_j} \geq j+2j (m_{j+1}+\cdots +  m_{k-1})\eq$$
Par exemple, le decompte des etats en $\A^{(1)}$ donne
$$\sum_{m_1\geq 0} z_1^{m_1} {q^{m_1^2+2m_1(m_2+\cdots m_{k-1} )}\over (q)_{m_1}}\eq$$

La condition de bord generale (qui remplace le nombre de 1 au bout qui peut etre  au plus $i-1$ dans l'expression d'Andrews) est:
$$ n^{(j)}_{m_j} \geq j+ {\rm max}\, (j-i+1,0) +2j (m_{j+1}+\cdots +  m_{k-1})\eq$$
soit une condition de bord pour chaque permier terme d'une sequence de $\A^{(j)}$. Ca ete teste dans des cas simples et ca reproduit la forme generale d'Andrews. Par exemple, pour $k=4$, on a
$$\A^{(1)}_{-n^{(1)}_1}\cdots \A^{(1)}_{-n^{(1)}_{m_1}} \A^{(2)}_{-n^{(2)}_1} \cdots\A^{(2)}_{-n^{(2)}_{m_2}}\A^{(3)}_{-n^{(3)}_1} \cdots\A^{(3)}_{-n^{(3)}_{m_3}} \eq$$avec
$$n^{(1)}_i \geq n^{(1)}_{i+1} + 2 \qquad{\rm et }\qquad   n^{(1)}_{m_1} \geq 1+2 (m_2+m_3)\eq$$
et
$$n^{(2)}_i \geq n^{(2)}_{i+1} + 4 \qquad{\rm et }\qquad   n^{(2)}_{m_2} \geq 2+4m_3\eq$$
et
$$n^{(3)}_i \geq n^{(3)}_{i+1} + 6\qquad{\rm et }\qquad   n^{(3)}_{m_3} \geq 3\eq$$
Ceci correspond au cas sans restriction - soit le cas ou on agit sur le vide ou sur le champs primaire $\s_1$. Si on agit sur $\s_2$, on a la condition de bord suivante:
$$n^{(3)}_{m_3} \geq 4 \eq$$ sans modification sur les indices $n^{(1)}_{m_1}$ et $n^{(2)}_{m_2}$ tandis que si on agit sur $\s_3$, on requiert
$$  n^{(2)}_{m_2} \geq 3+4m_3 \qquad {\rm et} \qquad n^{(3)}_{m_3} \geq 5\eq$$
sans changement sur $n^{(1)}_{m_1}$.

Pour le cas grade, la base sera la meme avec l'ajout a gauche d'une sequnece de modes $\A^{(1/2)}_{-n_i^{(0)}}$ avec 
$$n_i^{(0)}\geq n_{i+1}^{(0)} +1\eq$$ mais sans decallage du a la presence des autres modes a droite.

vskip0.3cm

%\vfill\eject
\centerline{\bf REFERENCES}
%\vskip 0.5cm
\immediate\closeout\refs \vskip 0.5cm
   \message{References}\input references
%\vfill\eject

\end